\documentclass[twocolumn,prb,showpacs,preprintnumbers,floatfix]{revtex4}

\usepackage[english]{babel}
\usepackage{graphicx}
\usepackage{rotating}
\usepackage{array}
\usepackage{amsmath}
\usepackage{amssymb}
\usepackage{amsfonts}
\usepackage{amsbsy}
\usepackage{mathbbol}

\renewcommand{\r}{\mathbf{r}}
\newcommand{\R}{\mathbf{R}}
\renewcommand{\k}{\mathbf{k}}
\newcommand{\q}{\mathbf{q}}

\newcommand{\be}{\begin{equation}}
\newcommand{\ee}{\end{equation}}
\newcommand{\abs}[1]{\left |{#1}\right |}
\newcommand{\ket}[1]{\left |\left.{#1}\right.\right\rangle}
\newcommand{\bra}[1]{\left\langle \left.{#1}\right.\right |}

\newcommand{\Eq}[1]{Eq.\,(\ref{#1})}
\newcommand{\Fig}[1]{Fig.\,\ref{#1}}
\newcommand{\Sec}[1]{Sec.\,\ref{#1}}
\newcommand{\Quot}[1]{Ref.\,\onlinecite{#1}}
\newcommand{\bwt}{\begin{widetext}}
\newcommand{\ewt}{\end{widetext}}

\begin{document}

\pagestyle{plain}

\title{Analysis of the exciton-exciton 
interaction in semiconductor quantum wells}

\author{Christoph Schindler}
\email{Christoph.Schindler@wsi.tum.de}
\author{Roland Zimmermann}
\affiliation{Humboldt-Universit\"at zu Berlin, Institut f\"ur Physik\\
Newtonstra{\ss}e 15, 12489 Berlin, Germany}

\begin{abstract}
The exciton-exciton interaction is investigated for quasi-
two-dimensional quantum structures. A bosonization scheme is
applied including the full spin structure. For generating the
effective interaction potentials, the Hartree-Fock and
Heitler-London approaches are improved by a full two-exciton
calculation which includes the van der Waals effect. With these
potentials the biexciton formation in bilayer systems is
investigated. For coupled quantum wells the two-body scattering
matrix is calculated and employed to give a modified relation
between exciton density and blueshift. Such a relation is of
central importance for gauging exciton densities in experiments
which pave the way toward Bose-Einstein condensation of
excitons.
\end{abstract}


\maketitle
\section{Introduction}\label{sec:intro}
Excitons (bound pairs of electron and hole) play a key role in
semiconductor optics, and their theoretical understanding has a
long history.\cite{Knox1} Although excitons obey Bose statistics,
the fermionic structure of their constituents is always
important and forbids to treat excitons as nearly ideal Bose
particles. Among the first attempts for bosonization of excitons, 
we quote Refs.\,\onlinecite{Usui1,Keldysh1,Haug1}. Both the
fermionic exchange and the Coulomb forces between two excitons
can be condensed into an effective exciton-exciton (XX)
interaction potential. Efforts for deriving XX potentials are
abound in the literature, covering three-dimensional excitons in
bulk semiconductors as well as quasi-two-dimensional excitons in
quantum wells. However, almost all of these
attempts\cite{Savona1,Brandes1,Rochat,Laikhtman1} were restricted
to the Hartree-Fock level, i.e., taking into account the XX
interactions up to second order in the elementary charge, $e^2$.
Sometimes it has been claimed that this would be enough for
treating sufficiently accurate the exciton gas at low density
$n_X$. This is not correct since already at zero density, the
single-exciton bound state calls for an infinite summation in
powers of the Coulomb potential, and consequently the next term
(linear in $n_X$) cannot be truncated either. It was not before
2001 that in a seminal paper by Okumura and Ogawa\cite{Okumura1}
the first XX potential beyond Hartree-Fock\cite{comment} for bulk
semiconductors has been constructed, in close analogy to the
Heitler-London approximation in atomic physics.

The interest in a proper description of XX interactions has been
surely intensified by the actual search for Bose-Einstein
condensation of excitons, which has been predicted theoretically
already decades ago.\cite{Moskalenko1,Keldysh2,Lozovik1} Due to the small
exciton mass compared to atomic systems, the critical temperature
for the condensate is expected to be a few kelvins for a density 
of $10^{17}\,\text{cm}^{-3}$ in bulk GaAs, within easy
reach experimentally. A fundamental problem, however, is the
finite life time of the excitons, which hinders the relaxation
into thermal equilibrium. One possible way out are coupled
quantum wells (CQWs) which came into focus
a few years ago.\cite{Butov5,Timofeev, Butov1,Snoke2,Snoke5} A static
electric field in growth direction forces electrons and holes to
reside in adjacent quantum wells which are separated by a
barrier. Due to this spatial separation, the indirect excitons
exhibit extremely long life times, which is a good condition for
reaching thermal equilibrium. However, these spatially indirect
excitons form dipoles leading to a strong and long-range
repulsion, which complicates the theoretical description as well as
the experimental realization of a dense cold exciton gas.
\cite{prb73_033319,prb74_045309}
Recently large progress has been reported for spatially indirect
excitons in electrostatic\cite{prl97_016803} as well as optical traps.
\cite{prl96_227402,prb76_193308}

These practical demands led us to investigate the XX interaction
in more detail, with special emphasis on coupled quantum wells.
In particular, we improve on the Heitler-London-type treatment in
\Quot{Okumura1} by solving the four-particle
Schr\"odinger equation for two electrons and two holes
numerically. Before doing so we address the complex spin
structure of the exciton composed of a spin 1/2 electron and a
spin 3/2 heavy hole (\Sec{sec:hamilton}). The importance of
spin-dependent effects in the exciton gas has been emphasized
e.g., in \Quot{Vina1}. In the limit of immobile holes
(their mass being usually much larger than that of the
electrons), we can derive effective spin-dependent XX potentials
(\Sec{sec:pot}) which contain in addition to the dipole-dipole
repulsion and exchange effects the weak van der Waals forces. The
latter is of importance to rectify a recent
claim\cite{Sugakov1,Wu1} to have explained the bead pattern
formation which appears at low temperatures in the ring-shaped 
CQW emission.\cite{Butov4,Butov2,Snoke3}

With a proper XX interaction at hand, we are able to calculate
biexciton states (excitonic molecules), which are well known from
bulk semiconductors and have been observed in single quantum
wells.\cite{Kleinman1,Kleinman2} Since biexciton energies have
been calculated with high precision elsewhere,\cite{Usukura1} we
use these results to judge our approximate treatment in
\Sec{sec:biexciton}. For the CQW situation simplified to a
bilayer system, we identify the parameter values (essentially
mass ratio and charge separation) which limit the existence of
biexcitons.

Addressing the many-exciton case, we will treat the excitons as
effective bosons with a renormalized interaction potential,
derived from the underlying electron-hole description.
\cite{Zim4} As an application, we investigate in
\Sec{sec:scattering} two realistic CQW structures. The
numerically generated XX potentials are used to calculate
two-exciton scattering phase shifts which are the main ingredient
for a $T$-matrix based quasi-particle dispersion\cite{Zim1}. For
the low-density case, we are able to calculate excitonic blueshift 
and scattering-induced broadening linear in $n_X$
(\Sec{sec:Xblue}). We find a stunning reduction of the blueshift
compared to the simple "capacitor formula" and relate this
finding to features in the XX pair-correlation function which -
due to the strong repulsion - resembles more a Fermi gas than a
free Bose gas (\Sec{sec:correlation}).

Conclusions are drawn in \Sec{sec:conclusions}, while a few
technical details are deferred to the Appendix.

\section{Many-Exciton Hamiltonian}
\label{sec:hamilton} To derive the many-exciton Hamiltonian we
follow the work of de-Leon and Laikhtman.\cite{Laikhtman1,Laikhtman5}
The task is to find the matrix elements of the electron-hole
Hamiltonian with an appropriate two-exciton wave function and to
implement them into an effective bosonic Hamiltonian for many
excitons. The effective-mass Hamiltonian for two electrons and
holes reads
\begin{multline} \label{eq:2ehH}
H_{2eh}=-\frac{\hbar^2\Delta_{e1}}{2m_e}-\frac{\hbar^2\Delta_{h1}}{2m_{h}}-
\frac{\hbar^2\Delta_{e2}}{2m_e}-\frac{\hbar^2\Delta_{h2}}{2m_{h}}\\
+v_{ee}(\r_{e1}-\r_{e2})+v_{hh}(\r_{h1}-\r_{h2})-v_{eh}(\r_{e1}-\r_{h1})\\
-v_{eh}(\r_{e2}-\r_{h2})-v_{eh}(\r_{e1}-\r_{h2})-v_{eh}(\r_{e2}-\r_{h1}),
\end{multline}
where the interaction part [second and third line of \Eq{eq:2ehH}] is composed
of Coulomb interactions between particle $a$ and $b$, which
can be either electron ($a=e$) or hole ($a=h$),
\be \label{eq:Coulomb3D} v_{ab}(\r)=\frac{e_0^2}{r} \quad
\text{with} \quad e_0^2=\frac{e^2}{4\pi\epsilon_0\epsilon_s}. \ee
Please note that the corresponding sign (attractive or repulsive) 
has been made explicit in \Eq{eq:2ehH}.

We define exciton quantum field operators $\Psi_s^{\dagger}(\R)$
which create an exciton at exciton center-of-mass (c.m.)
position $\R$ with spin $s$, and $\Psi_s(\R)$, which annihilates
the same exciton. The subsequent Hamilton operator will be
written as
\bwt
\be \label{eq:ManyXHam}
H_{XX}=\int d\R\sum_{s}\Psi^{\dagger}_s(\R)\frac{-\hbar^2
\Delta_{\R}}{2M}\Psi_s(\R)+\frac{1}{2}\int d\R d\R' 
\sum_{s1s2s3s4}W_{s1s2,s3s4}(\R-\R')\Psi_{s1}^{\dagger}(\R)
\Psi_{s2}^{\dagger}(\R')\Psi_{s3}(\R')\Psi_{s4}(\R),
\ee
\ewt
with the exciton mass $M=m_e+m_h$ in the kinetic energy. $W_{s1s2,s3s4}(\R)$
is the spin and space-dependent pair interaction potential. In the
following, it will be extracted from a careful study of the four-particle
problem [\Eq{eq:2ehH}]. The exciton spin index $s=s_e+J_h$ 
is the sum of the electron spin
($s_e=\pm 1/2$) and the heavy-hole angular momentum ($J_h=\pm
3/2$). We neglect the light holes, which are separated due to the 
confinement effects in the quasi-two-dimensional quantum well
(QW). Now, $s$ runs over four values $s=\pm1$ (bright states) and
$s=\pm2$ (dark states). Since electron and hole are fermions, we
have to use a properly symmetrized ansatz for both component's
spin wave functions $\chi^{p}_{ee}(s_{e1},s_{e2})$ and
$\chi^{q}_{hh}(J_{h1},J_{h2})$, which make the overall wave-
function antisymmetric with respect to the exchange of the
electrons and the holes, respectively. The labels $p$ and $q$ denote
the parity which can be symmetric $(p=s)$ or antisymmetric
$(p=a)$. Together with the spatial part we write the total
two-exciton wave function,
\begin{multline}
\Psi^p_q(\r_{e1},s_{e1},\r_{e2},s_{e2},\r_{h1},J_{h1},\r_{h2},J_{h2})\\
=\psi^{p}_{q}(\r_{e1},\r_{e2},\r_{h1},\r_{h2})\chi^{-p}_{ee}(s_{e1},s_{e2})
\chi^{-q}_{hh}(J_{h1},J_{h2}),
\end{multline}
where the upper label stands for the electron part, and the lower
one for the hole part. Please note that due to the fermionic
nature of the particles, the parities of the spin part and of the
spatial part have to be opposite to each other. Taking now the
matrix elements of \Eq{eq:2ehH} with the ground state wave
functions we get four different potentials,\cite{Quattropani1}
\be \label{eq:Utt}
U^{p}_{q}=\bra{\Psi^p_q}H_{2eh}^{\text{int}}\ket{\Psi^p_q}.
\ee
Due to the one to one correspondence between the exciton spin and
the spin of its constituents, it is possible to express the basis
vectors in the space of symmetric and antisymmetric spin wave
functions by the spin eigenstates of the excitons
$\ket{s_1,s_2}$. After a straightforward unitary transformation
one gets the interaction matrix elements $W_{s1s2,s3s4}(\R)$ as
shown in Table \ref{tab:tbl1_W_spin_struct} for columns $\ket{s_3s_4}$ 
and rows $\bra{s_1s_2}$. Here
$U^{\pm}_{q}=1/2\left( U^a_{q}\pm U^s_{q}\right)$ and similar for
the lower hole index. 
\begin{table}[ht]
    \includegraphics{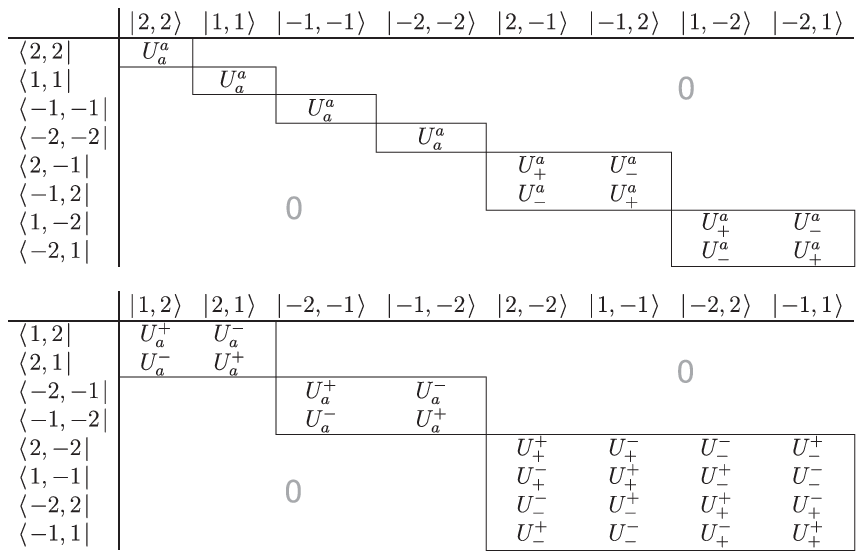}
    \caption{Spin structure of the exciton-exciton interaction 
    $W_{s1s2,s3s4}(\R)$ in quantum wells.}
    \label{tab:tbl1_W_spin_struct}
\end{table}
\begin{table}[ht]
    \includegraphics{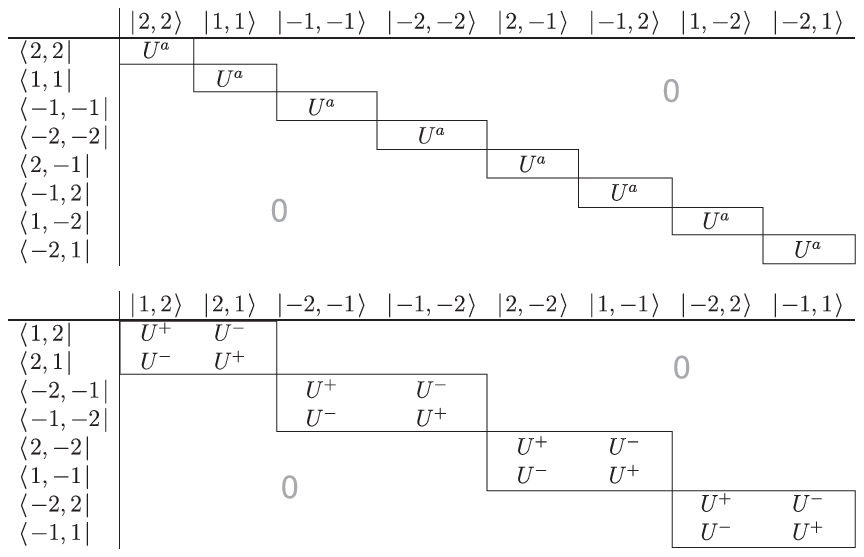}
    \caption{Spin structure of the exciton-exciton interaction 
    $W_{s1s2,s3s4}(\R)$ in the case of infinitely heavy holes.}
    \label{tab:tbl2_W_0_spin_struct}
\end{table}
We can clearly see the block structure of the interaction which
reflects the conservation of the total spin $s_1+s_2=s_3+s_4$.
The interaction channels can be classified as follows: There is
the direct channel where an initial state $\ket{s_1,s_2}$ will
remain unchanged. The other channel is of exchange type with a
change in the initial state due to three different processes:
Exchange of electrons, exchange of holes, and exchange of both
simultaneously. The electron-hole exchange process
(longitudinal-transverse splitting of the exciton) has been
neglected already in the starting Hamiltonian.

In a case where the hole mass is much larger than the electron
mass, the exchange of holes becomes negligible and the four
different potentials [\Eq{eq:Utt}] collapse into two, $U^a=U^a_{q}$
and $U^s=U^s_{q}$, which correspond to wave functions properly
symmetrized only with respect to the electrons. In this case the
interaction part of the Hamiltonian simplifies to the form shown
in Table \ref{tab:tbl2_W_0_spin_struct} where $U^{\pm}=1/2\left(
U^a\pm U^s\right)$ and can be cast into the form given in the
Appendix, [\Eq{eq:ManyXHamInt}].

\section{Effective interaction potentials}
\label{sec:pot}
To derive the spatial dependence of the interaction potentials
introduced in \Sec{sec:hamilton}, approximations have been
introduced in the literature. In the infinitely heavy-hole limit,
the Heitler-London ansatz is well known from atomic physics and
has recently been brought into exciton physics by Okumura and
Ogawa.\cite{Okumura1} In a bulk semiconductor, the two excitons
resemble a hydrogen molecule in this limit. The problem of four
particles (two electrons and two holes) simplifies here to a
two-particle problem for the electrons, while the hole-hole
distance $\R$ enters as a parameter. Therefore, the Coulomb
potential between the holes gives just a fixed additional term in
the Hamiltonian. The two-exciton wave function can then be
written as a properly antisymmetrized product of single-exciton
wave functions in the $1s$ ground state $\phi(\r)$, centered
around the position of each hole. For the spatial part of the
two-exciton wave function, we can write 
\be \label{eq:HLwf}
\psi^{s,a}(\r_1,\r_2) \,=
\,\frac{1}{\sqrt{2}}\,\frac{\phi(\r_1)\,\phi(\r_2-\R)
\,\pm\,\phi(\r_1-\R)\, \phi(\r_2)}{\sqrt{1 \, \pm
\,\mathcal{O}^2(\R)}}, \ee 
with the wave function overlap, 
\be
\mathcal{O}(\R) = \int d\r \,\phi(\r)\,\phi(\r-\R), 
\ee 
which leads to the potentials, 
\be \label{eq:UsUt}
\begin{split}
U^s(\R) = \frac{U_d(\R) - U_x(\R)}{1+\mathcal{O}^2(\R)}, \\
U^a(\R) = \frac{U_d(\R) + U_x(\R)}{1-\mathcal{O}^2(\R)},
\end{split}
\ee 
with direct $U_d(\R)$ and exchange $U_x(\R)$ potential
[explicit expressions are given in the Appendix, \Eq{eq:Ud} and
\Eq{eq:Ux}].

A simplified version of the Heitler-London approximation, which
has usually been applied to excitonic
systems,\cite{Savona1,Zim2,Brandes1,Laikhtman1} is the famous
Hartree-Fock treatment where the normalization denominator in
\Eq{eq:HLwf} is left out and is consequently missing in
\Eq{eq:UsUt}. This scheme can easily be generalized to arbitrary
hole masses. However, it leads to a nonlocal exchange
potential\cite{Savona1} and to problems with the orthogonality of
the basis states.\cite{Laikhtman6}

\begin{figure}[ht]
    \includegraphics{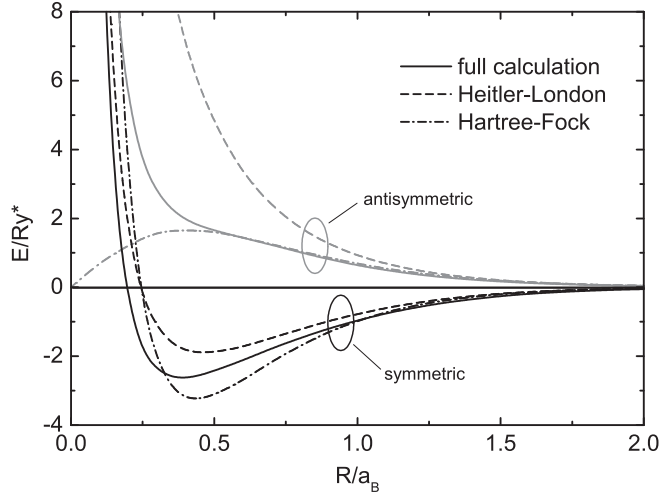}
    \caption{Effective interaction potentials for the strictly 
    two-dimensional system: full calculation results (solid lines), 
    Heitler-London approximation (dashed lines), and Hartree-Fock 
    approximation (dot-dashed lines) plotted vs distance in units of the 
    three-dimensional (3D)
    (bulk) exciton Bohr radius $a_B$. The vertical axis is given in units 
    of the bulk exciton Rydberg energy $Ry^*$. Antisymmetric channels 
    are displayed in gray, while symmetric channels have black lines.}
    \label{fig:fig1_potentials_strict2D}
\end{figure}

To improve over the approximations discussed so far, we have
solved the Schr\"odinger equation for the two electrons
numerically using the Lanczos algorithm, treating the holes as
infinitely heavy and thus immobile. The equation to be solved
reads
\be
\begin{split}
\label{eq:SchroedingerLanczos}
\left[-\frac{\hbar^2}{2\mu}\right.\left(\Delta_{\r_1}+ \Delta_{\r_2}\right)+
v_{ee}(\r_2-\r_1)+v_{hh}(\R)-v_{eh}(\r_1) \\
-v_{eh}(\r_2)-v_{eh}(\r_2-\R)-v_{eh}(\r_1-\R)\biggr] \Psi^{p}(\r_1,\r_2) \\
=\left(-2B_{X} + U^{p}(\R)\right)\Psi^{p}(\r_1,\r_2),
\end{split}
\ee 
where $\mu$ is the effective electron mass in the plane and
$B_X$ is the single-exciton binding energy. The zero of energy is
chosen to be the band gap. \Fig{fig:fig1_potentials_strict2D} 
shows the resulting effective interaction potentials in 
Hartree-Fock, Heitler-London, and full numerical quality 
exemplarily for a strictly two-dimensional system. We note the 
unphysical behavior of the Hartree-Fock antisymmetric potential 
which approaches zero for small distances. It also misses the 
proper sequence of the antisymmetric channel to be above the 
symmetric one for small distances. This is the result of leaving 
out the normalization denominator in \Eq{eq:HLwf}. Thus these 
problems are corrected in the Heitler-London treatment. 
Here also the antisymmetric channel shows the expected Coulomb 
singularity of the hole-hole potential for small distances. In 
the full solution the energy is lowered
compared to the Heitler-London approximation by a mutual
deformation of the excitonic orbitals. Please note that the van
der Waals effect is included due to the nonperturbative nature
of the calculation.

While the considerations above are correct also in three
dimensions, we turn now to the two-dimensional case of quantum
structures. All vectors are to be understood as lying in the
x-y plane, the z-axis being the growth direction.

\begin{figure}[ht]
    \centering
       \includegraphics{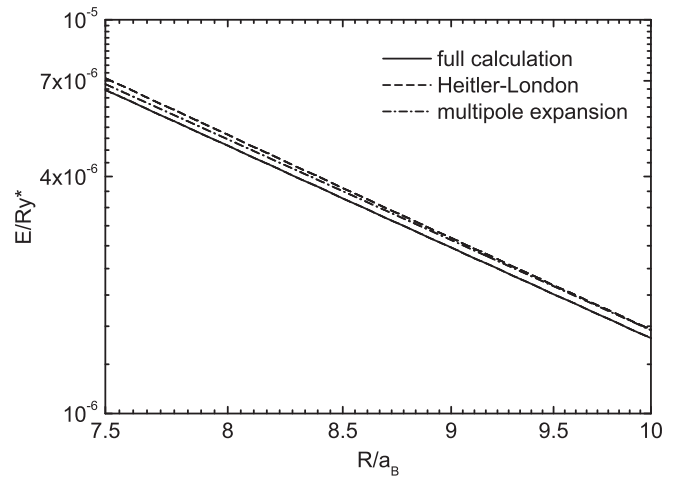}
    \caption{Asymptotic behavior of the potentials on a double 
    logarithmic plot for the strictly two-dimensional system: 
    full calculation (solid), Heitler-London (dashed) and multipole
    expansion (dot-dashed); symmetric states lie on top of antisymmetric ones.}
    \label{fig:fig2_2D_asym1}
\end{figure}

\begin{figure}[ht]
    \centering
        \includegraphics{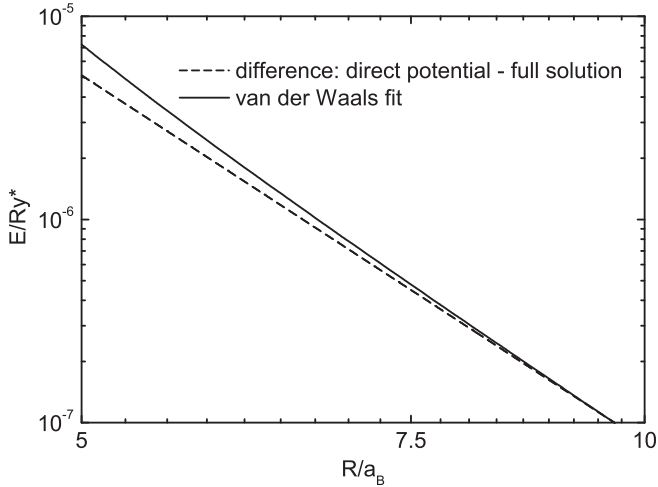}
    \caption{The van der Waals effect in the strictly two-dimensional 
    system. Deviation of the full calculation from the direct potential 
    for asymptotic large distances (solid line) and fitted van der Waals 
    like potential $\propto 1/R^6$ (dashed line).}
    \label{fig:fig3_2D_asym2}
\end{figure}

We note that all interaction potentials in \Fig{fig:fig1_potentials_strict2D}
approach zero from positive values and hence show a repulsion for
larger distances, although this is hardly seen in 
\Fig{fig:fig1_potentials_strict2D}. However, this behavior follows 
already from a multipole expansion
of the direct potential $U_d(\R)$, which dominates the
interaction at large distances. In such an approach we treat the
exciton as a static charge distribution $\rho(\r,z)$ with
cylindrical symmetry and centered at zero c.m. coordinate. The
specific form of $\rho(\r,z)$ depends on the system under
investigation and will be specified later. Due to the charge
neutrality of the exciton the multipole expansion starts with the
dipole term $\propto 1/R^3$, where $\R$ denotes the in plane
c.m. distance of the two excitons. Up to the quadrupole term we
obtain for the asymptotics
\begin{multline} \label{eq:PoEx2}
R\rightarrow\infty:\, U_d(\R) = \frac{e_0^2}{R^3}\left\langle
 z\right\rangle ^2\\
 + \frac{e_0^2}{R^5}\left( \frac{9}{4}\left\langle x^2-z^2
 \right\rangle ^2+3\left\langle z\right\rangle\left\langle z 
 \left(3x^2 - z^2\right)\right\rangle\right),
\end{multline}
where the angular brackets denote averaging over $\rho(\r,z)$.
For a bulk system with spherical symmetry we see immediately that
the multipole expansion vanishes as expected. However, for the
reduced symmetry of quasi-two-dimensional systems, we get finite
multipoles also for in-plane circular symmetry. Even for the
strictly two-dimensional system, where $z\equiv0$, there is a
contribution from the quadrupole-quadrupole interaction ($\propto
1/R^5$). This holds as well for symmetric single QWs where
$\left\langle z\right\rangle=0$ due to the mirror symmetry along
the z-axis. The direct interaction thus falls off as a power law
rather than exponentially, which has been overlooked in a recent
investigation of the asymptotic XX potential.\cite{Sugakov1} In
\Fig{fig:fig2_2D_asym1} we compare the asymptotic behavior of the
numerically calculated direct potential [\Eq{eq:Ud}], and the
result from the multipole expansion [\Eq{eq:PoEx2}] using the
strictly two-dimensional charge distribution for an immobile hole
$\rho(\r,z)=\delta(z)\left[\delta(\r)-\phi^2(\r)\right]$ and the
usual Coulomb potentials [\Eq{eq:Coulomb3D}]. The multipole
expansion holds for $R\gtrsim 9\,a_B$. The full solution shows
the van der Waals effect and lies below the direct potential.
However, this effect is not able to overcome the repulsive nature
of the asymptotic potential.
To grasp this effect a bit more quantitatively, we plot in
\Fig{fig:fig3_2D_asym2} the difference between the direct potential
from the approximations of rigid orbitals and a corresponding
quantity for the full solution
($U_d^{\text{full}}=1/2(U^a+U^s)$). For very large distances
$R\ge 10\,a_B$, we fit this difference to a van der Waals
potential, i.e., a $\propto 1/R^6$ power law (dashed line in
\Fig{fig:fig3_2D_asym2}). The van der Waals law holds only for
distances $R\gtrsim 9\,a_B$.
For smaller distances higher-order effects come in and spoil the 
$\propto 1/R^6$ dependence.

With the derived effective interaction potentials, we construct
the two-dimensional Schr\"odinger equation for two excitons with
mutual distance $R$ in c.m. space, where we introduce now the
finite exciton mass $M=m_e+m_h$ in the kinetic term:
\begin{multline} \label{eq:RadialSchroedingerBxx}
\left\{\frac{\hbar^2}{M}\left( -\frac{1}{R}\frac{d}{dR}R\frac{d}{dR}+
\frac{m^2}{R^2}\right)+U^{p}(R)\right\} \psi^{p}_m(R)\\=E_m^{p}\psi_m^{p}(R),
\end{multline}
where $m$ denotes the angular quantum number. Please note the missing
factor of 2 in the denominator of the kinetic term since we
have to consider the reduced mass of two excitons, $\mu_X=M/2$.

\section{Model Systems}
\label{sec:biexciton} 
\begin{figure}[ht]
    \includegraphics{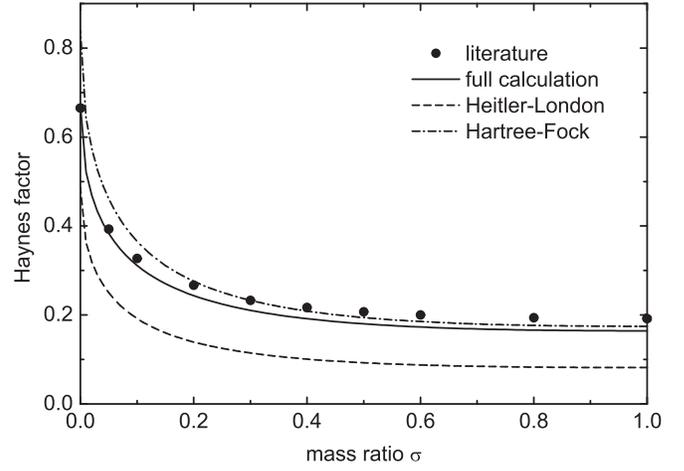}
    \caption{Haynes factor $f_H=B_{XX}/B_{X}$ for the different potentials 
    in the strictly two-dimensional system: full calculation (solid line), 
    Heitler-London potential (dashed line), and Hartree-Fock potential 
    (dot-dashed line) plotted vs mass ratio $\sigma=m_e/m_h$. The
    dots are numerically exact values from Ref.\,\onlinecite{Usukura1}.}
    \label{fig:fig4_Haynes_massratio_comp}
\end{figure}

To test the reliability of our
Born-Oppenheimer-type method, we calculate the biexciton binding
energy $B_{XX}$ from the lowest state of
\Eq{eq:RadialSchroedingerBxx} with $m=0$ for different mass
ratios $\sigma=m_e/m_h$ in the strictly two-dimensional limit.
These results are compared in \Fig{fig:fig4_Haynes_massratio_comp} with
variational calculations from the literature\cite{Usukura1} which
are numerically exact. As expected, our method produces exact
results for $\sigma=0$. For nonzero mass ratio we underestimate
the binding energy slightly, e.g., for $\sigma=0.3$ we have an
error of $\approx8\%$. Even for $\sigma=1$ the error is only
about $12\%$. We conclude that the method is a reasonable
approximation for GaAs quantum structures ($\sigma=0.3$), which
will be under investigation in \Sec{sec:scattering}.
We plot the Haynes factor $f_H=B_{XX}/B_{X}$ for the other
approximations as well (for the strictly two-dimensional system,
$B_X = 4 Ry^*$). The Heitler-London approximation underestimates
the biexciton binding energy significantly. Hartree-Fock seems to
do much better, but note that for small $\sigma$ the Haynes
factor is overestimated, which is in contrast to the variational
principle.

\begin{figure}[ht]
     \includegraphics{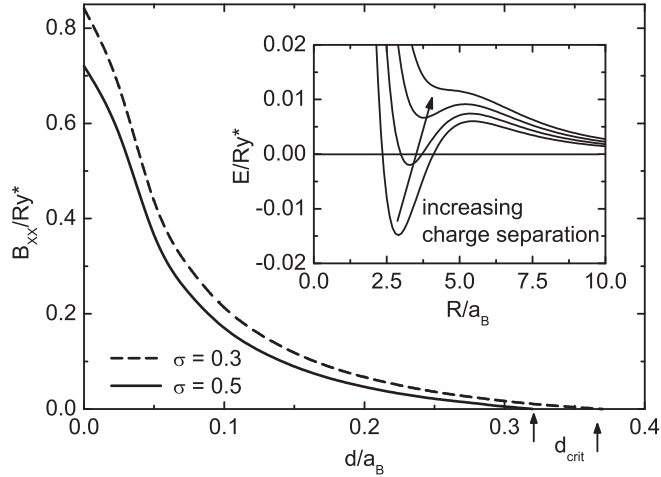}
    \caption{Biexciton binding energy in the bilayer system plotted 
    vs charge separation at $\sigma=0.5$ (solid line) and $\sigma=0.3$ 
    (dashed line). The arrows denote the corresponding critical charge 
    separation $d_{\text{crit}}$. Curves in the inset show the symmetric 
    potential $U^s(R)$ for $d=0.8\text{, }0.9\text{, }1.0\text{, and }1.1$ 
    in units of the 3D exciton Bohr radius $a_B$.}
    \label{fig:fig5_Bxx_cs_mass_05_03}
\end{figure}

\begin{figure}[ht]
    \includegraphics{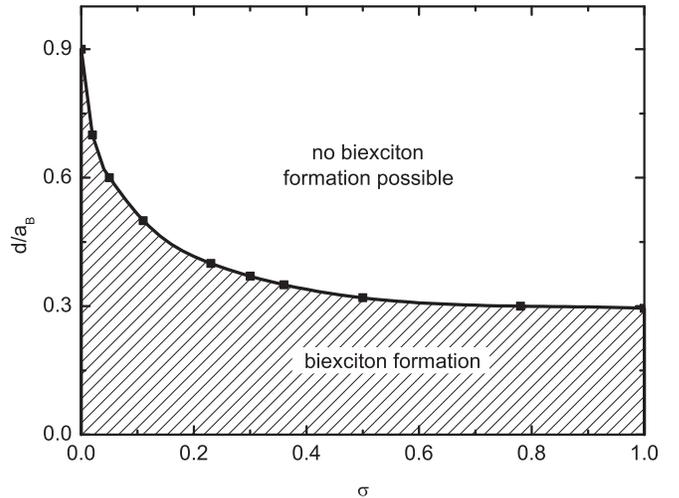}
    \caption{Critical charge separation in units of the bulk exciton 
    Bohr radius $a_B$ plotted vs mass ratio $\sigma=m_e/m_h$.}
    \label{fig:fig6_phase_diagram_bilayer}
\end{figure}

A simple model for a coupled quantum well structure is the so-
called bilayer system:\cite{Tan1} Electrons and holes are
confined each in infinitely narrow planes with a separation $d$
between the layers. This separation of unequal charges leads to a
reduced Coulomb interaction between particles in different layers
but leaves the potentials between particles of the same kind
unchanged,
\be \label{eq:bilayer} v_{eh}(\r) =
\frac{e_0^2}{\sqrt{r^2+d^2}}\,,\quad
v_{ee}(\r)=v_{hh}(\r)=\frac{e_0^2}{r}. \ee
The charge distribution reads now
$\rho(\r,z)=\delta(z)\delta(\r)-\delta(z-d)\phi^2(\r)$, and the
expansion [\Eq{eq:PoEx2}] yields a finite dipole-dipole interaction
$\propto 1/R^3$, resulting in a strong long-range repulsion. For
the many-exciton problem we are in particular interested in the
biexciton formation. \Fig{fig:fig5_Bxx_cs_mass_05_03} shows the 
biexciton binding energy versus charge separation for a mass ratio 
of $\sigma=0.3$ and $\sigma=0.5$. In both cases we find a fast 
reduction of $B_{XX}$ with increasing separation between the 
layers as expected. At a certain critical charge separation 
$d_{\text{crit}}$, denoted by arrows in \Fig{fig:fig5_Bxx_cs_mass_05_03}, the 
biexciton ceases to exist. The same model system has been investigated by 
Tan et al.\cite{Tan1} using quantum Monte Carlo (QMC) technique.
From their results at intermediate values of $d$, 
which agree nicely with our own calculations, the authors 
suggested an exponential decay of $B_{XX}$ for large $d$. 
The inset of \Fig{fig:fig5_Bxx_cs_mass_05_03}
shows the potential in the symmetric channel for different values 
of the charge separation. It is clearly seen how the potential 
minimum passes through zero with increasing $d$ and vanishes 
completely if $d$ is enlarged further (large arrow). Therefore 
a (finite) critical charge separation $d_{\text{crit}}$ exists 
where the biexciton binding energy becomes zero due to the 
dominant dipole-dipole repulsion. More recently this critical behavior
was also observed in QMC calculations of Lee and Needs.\cite{privCommNeeds}
In \Fig{fig:fig6_phase_diagram_bilayer} we plot the critical charge separation
vs mass ratio. Below this curve, in the shaded area, bound
states can be formed, while above no biexcitons exist. For small
$\sigma$ we obtain a rapid decrease of the critical charge
separation. For larger mass ratios, on the other hand, the
separation does not depend much on $\sigma$. This behavior
resembles \Fig{fig:fig4_Haynes_massratio_comp}, revealing a direct connection
between biexciton binding energy and critical charge separation.

\section{Results for Coupled Quantum Wells}
\label{sec:scattering} 
\begin{figure}[ht]
  	\includegraphics{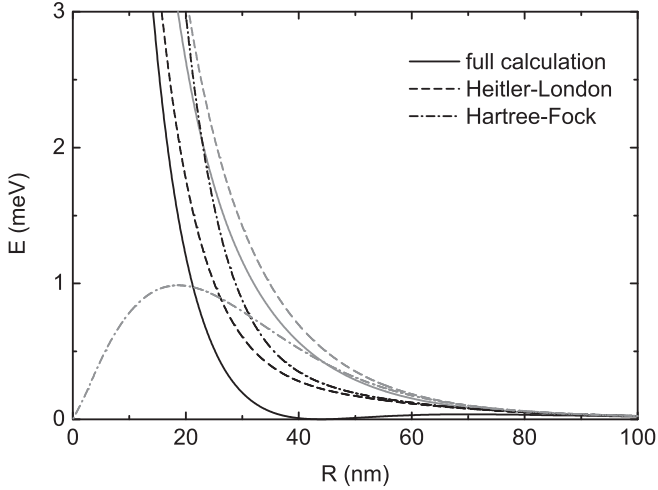}
    \caption{Interaction potentials for sample A: full calculation results 
    (solid lines), Heitler-London potentials (dashed lines), and 
    Hartree-Fock potentials (dot-dashed lines) plotted vs distance. 
    Antisymmetric channels have gray lines, while symmetric channels have 
    black ones.}
    \label{fig:fig7_potentials_Butov}
\end{figure}

\begin{figure}[ht]
    \includegraphics{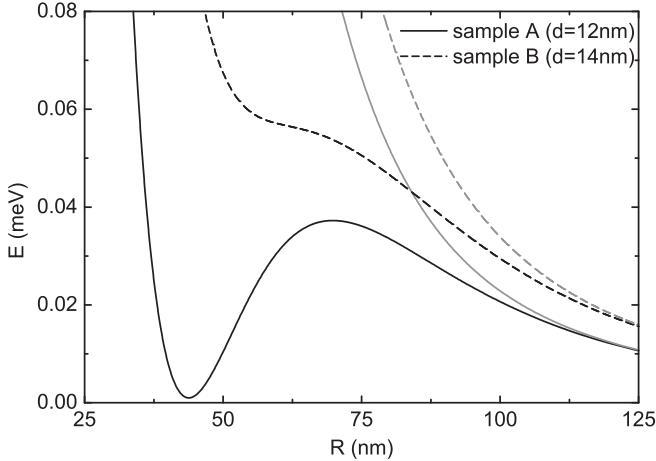}
    \caption{Comparison of the full calculation results for sample A 
    (solid lines) and sample B (dashed lines). Antisymmetric 
    channels have gray lines, while symmetric channels have black ones.}
    \label{fig:fig8_potentials_Snoke_Butov}
\end{figure}

We turn now to realistic coupled quantum
well structures, having GaAs as well material and barriers made
of $\text{Al}_x\text{Ga}_{1-x}\text{As}$. In this case we treat
the Coulomb interaction $v_{ab}(\r)$ in single sublevel
approximation,\cite{Laikhtman2,Zim4} which is given by
\be \label{eq:vab} v_{ab}(\r)=e_0^2\int dz' dz
\frac{u_a^2(z')u_b^2(z'-z)}{\sqrt{r^2+(z-z')^2}}. \ee
Here, $\r$ is again an in-plane vector, and $u_{e}(z)$ and
$u_h(z)$ denote the confinement functions of the lowest sublevel
for electron and hole. They enter as well the static charge
distribution of the exciton:
$\rho(\r,z)=u^2_h(z)\delta(\r)-u^2_e(z)\phi^2(\r)$.
\Fig{fig:fig7_potentials_Butov} shows the effective interaction potentials for
a GaAs/$\text{Al}_{0.3}\text{Ga}_{0.7}\text{As}$ coupled quantum
well geometry used by Butov\cite{Butov2} with a nominal (i.e.,
center distance between the wells) charge separation of
$d=12\,\text{nm}$ (sample A in  Table \ref{tab:CQWdetails}). We
see again the already discussed features of the three
approximation levels. The energy gain when going from
Heitler-London to the full calculation yields a minimum for the
symmetric channel, which however is so weak that no biexcitons
can be formed. This feature strongly depends on the geometry of
the quantum wells, which is illustrated in 
\Fig{fig:fig8_potentials_Snoke_Butov}.
Here the full calculation potentials for sample A are compared to
another one used by Snoke et al.\cite{Snoke1} where $d=14\,\text{nm}$
(sample B in Table \ref{tab:CQWdetails}). Due to the larger
charge separation no minimum can be seen in the symmetric
channel. This observation is consistent with the results obtained
for the idealized (bilayer) model discussed in
\Sec{sec:biexciton}.
\begin{table}[ht]
    \caption{Details of the used CQW geometry together with the
    calculated binding energy $B_X$ of the
    indirect exciton.}
    \label{tab:CQWdetails}
    \begin{tabular}{l c c}
    \hline \hline
    ~ & Sample A & Sample B\\ \hline
    Well width $L_z$ (nm) & $8.0$ & $10.0$ \\
    Barrier width (nm) & $4.0$ & $4.0$ \\
    $d$ (nm) & $12.0$ & $14.0$ \\
    $d_{\text{eff}}$ (nm) & $10.8$ & $12.7$ \\
    Static field (kV/cm) & 30.0 & 36.0 \\
    $B_X$ (meV) & $4.0$ & $3.5$\\
    \hline \hline
    \end{tabular}
\end{table}

With these results a simple explanation for the regular bead
pattern in the luminescence ring at low
temperatures\cite{Butov4,Butov2,Snoke3} has to be ruled out: It was
speculated\cite{Sugakov1,Wu1} that the van der Waals effect could
overcome the dipole-dipole repulsion, resulting in an attraction
between spatially indirect excitons, which would lead to a
spontaneous patterning. Our calculation shows that this is not
the case in agreement with recent experimental investigations.
\cite{prl92_117405,prb75_033311} In \Quot{Sugakov1} the 
quadrupole-quadrupole
interaction in two-dimensional systems has been overlooked and
hence the role of the van der Waals force overestimated as
discussed in \Sec{sec:biexciton}.

\section{Exciton Blueshift and Broadening}
\label{sec:Xblue} 
We turn now to the calculation of the
interaction induced blueshift and broadening which can be
observed in photoluminescence
experiments.\cite{Snoke1,Butov3,Snoke4} In a point charge
treatment of spatially indirect excitons, we have from \Eq{eq:Ud}
\be U_d(\R)=v_{hh}(\R)+v_{ee}(\R)-2v_{eh}(\R). \ee
Plugging in \Eq{eq:bilayer} for the bilayer system, one gets a
dipole-dipole repulsion of the form
\be \label{eq:PcDipoledipole} U_d(\R)\approx
e_0^2\left[\frac{2}{R}-\frac{2}{\sqrt{R^2+d^2}}\right]. \ee
Assuming a homogeneous exciton density $n_X$, this leads to a blueshift,
\be \label{eq:Capacitor} \Delta_0=\int d^2R\, U_d(\R)\,n_X =
d\frac{e^2}{\epsilon_0\epsilon_s}n_X. \ee
Since this expression is consistent with the electrostatics of a
plate capacitor, it is often referred to as capacitor
formula.\cite{Zhu1} We will derive a corrected formula for the
blueshift and the scattering-induced broadening using the
effective interaction potentials given in 
\Fig{fig:fig8_potentials_Snoke_Butov}. The
exciton self energy is calculated in a $T$-matrix
approach.\cite{Zim1} In the low-density limit and assuming
complete spin equilibrium, we write the two-body $T$-matrix equation
as
\be \label{eq:Tmatrix}
\bra{\q}T^{p}(z)\ket{\q''}=U^{p}_{\q-\q''}-\sum_{\q'}\frac{U^{p}_{\q-\q'}}
{2\epsilon_{\q'}-\hbar z}\bra{\q'} T^{p}(z)\ket{\q''}. \ee
The $T$-matrix enters the quasiparticle self-energy as
boson-direct (D) and boson-exchange (X),
\begin{multline} \label{eq:ScatAmp}
\Sigma_{\k}(\epsilon_{\k})=4\sum_{\q}\left\{
\bra{\q}T^D(z)\ket{\q}+\bra{\q}T^X(z)\ket{-\q}\right\}\\
\times n_B\left( \frac{\hbar^2}{2M}(\k+2\q)^2-\mu\right),
\end{multline}
where $\hbar z=\hbar^2 q^2/M+i0$ is put on shell and
$n_B(\epsilon)$ is the exciton
\begin{figure}[ht]
    \includegraphics{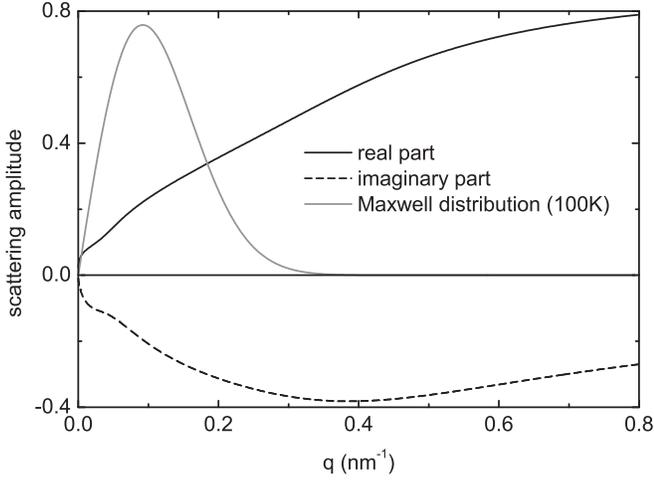}
    \caption{Real (solid line) and imaginary (dashed line) parts of the 
    scattering amplitude as a function of momentum for sample A. 
    Included is the Maxwell 
    distribution at $T=100\,\text{K}$ to show the relevant values 
    for $q$. The ordinate has been normalized to the capacitor value 
    [\Eq{eq:Capacitor}].}
    \label{fig:fig9_ReT_ImT_Butov}
\end{figure}
distribution function, which is later taken as the low-density
Maxwell-Boltzmann expression. The spin structure of the
Hamiltonian yields the following decomposition in the limit of
immobile holes:
\be T^{D}=3T^a+T^s \quad \text{and} \quad T^X=3/2\,T^a-1/2\,T^s.
\ee
The on-shell $T$-matrix needed in \Eq{eq:ScatAmp} depends
exclusively on the phase shifts $\delta^p_m(q)$
via\cite{Morgan1}
\be \label{eq:TpartialWave}
\bra{\q}T^{p}(z)\ket{\pm\q}=\frac{\hbar^2}{M}\sum_{m} (\pm
1)^m \frac{4}{i-\cot(\delta^{p}_m(q))}. \ee
We have extracted the phase shifts from the asymptotics of the
solution of the radial Schr\"odinger equation
[\Eq{eq:RadialSchroedingerBxx}] for $E=\hbar^2q^2/M$. Results for
the total complex scattering amplitude [curly bracket in
\Eq{eq:ScatAmp}] are shown in \Fig{fig:fig9_ReT_ImT_Butov}.
Please note that for large momenta $q$, the real part of the
scattering amplitude approaches the prediction of the capacitor
formula but with an effective charge separation $d_{\text{eff}}$
which is somewhat below the nominal one (third and forth row of
Table \ref{tab:CQWdetails}). This reduction is due to the spatial
extension of electron and hole charges along $z$. A simple
argument assuming confinement wave functions for infinite
barriers leads to $d_{\text{eff}} \approx d - 0.1267 L_z$.
This is quite close to the numerical value which follows from
replacing the point charge potential [\Eq{eq:PcDipoledipole}] in
\Eq{eq:Capacitor} by the numerically derived one.

\begin{figure}[ht]
    \includegraphics{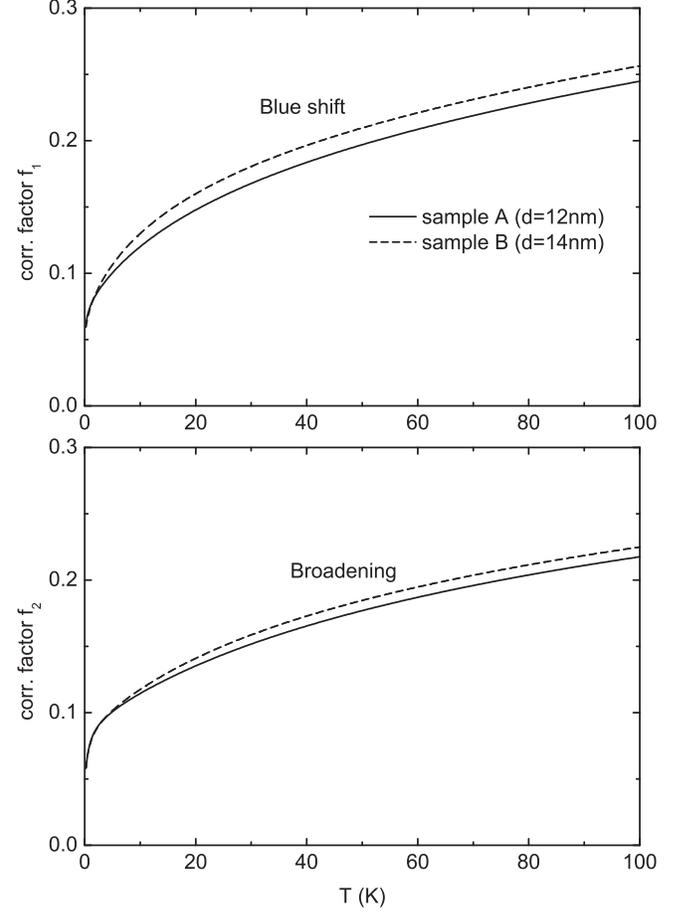}
    \caption{Correction factors to the capacitor formula in dependence on 
    temperature for sample A (solid lines) and sample B (dashed lines). 
    The upper panel shows $f_1$ (blueshift), while the lower one
    shows $f_2$ (broadening).}
    \label{fig:fig10_correction_Butov_Snoke}
\end{figure}

Using the scattering amplitude, we calculate the quasiparticle
shift and broadening at the dispersion edge ($\k=0$) and
introduce correction factors $f_1(T)$, $f_2(T)$ to the capacitor
formula
\be \label{eq:CapacitorCorr}
\Sigma_0(0)=d\frac{e^2}{\epsilon_0\epsilon_S}n_X\left(f_1(T)-if_2(T)\right).
\ee
The real part of this quantity is the blueshift $\Delta$ of the
exciton due to the repulsive interaction, while the imaginary
part can be associated with a finite broadening. Please note that
with the sign convention used, $\text{Im}\Sigma$ is negative. The
correction factors shown in \Fig{fig:fig10_correction_Butov_Snoke} reduce
the capacitor result dramatically. Therefore, the density for a
measured blueshift would be underestimated by a factor of 10 at
low temperatures. The broadening is of the same order of
magnitude, which is consistent with experimental
findings.

\section{Exciton-Exciton correlation function}
\label{sec:correlation} 
\begin{figure}[ht]
    \includegraphics{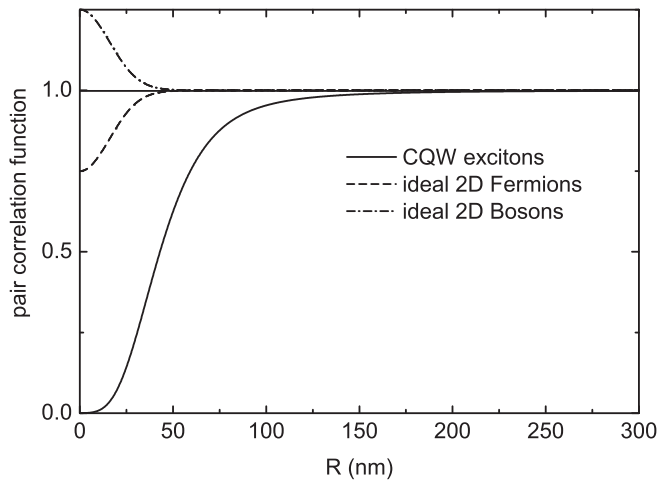}
    \caption{Exciton pair-correlation function vs distance at a 
    temperature of $T=6\,\text{K}$. The solid line refers to the 
    calculated exciton-exciton potential for sample A. The dashed
    line represents ideal bosons, while the dot-dashed line holds for 
    ideal fermions.}
    \label{fig:fig11_Pair_correlation_Butov}
\end{figure}

The significant reduction of the
quasiparticle shift compared to the capacitor value can be
explained with a strong depletion of the exciton gas around a
given exciton due to the repulsive interaction. To grasp this
repulsive correlation more directly we calculate the exciton pair-
correlation function:
\be \begin{split}
g_{ss'}(\R)&=\frac{\left\langle \Psi_s^{\dagger}(\R)\Psi_{s'}^{\dagger}(0)
 \Psi_{s'}(0)\Psi_s(\R)\right\rangle}{\left\langle\Psi_s^{\dagger}(\R)
 \Psi_s(\R) \right\rangle\left\langle\Psi_{s'}^{\dagger}(0)\Psi_{s'}(0) 
 \right\rangle}\\
&=\frac{1}{n_sn_{s'}}\sum_{\k\k'\q}e^{i\q\R}\left\langle \Psi_{\k s}^{\dagger}
\Psi_{\k'+\q s'}^{\dagger}\Psi_{\k' s'}\Psi_{\k+\q s}\right\rangle .
\end{split}
\ee
It has the same spin structure as the $T$-matrix. In the spin
equilibrated situation investigated here, the exciton density
$n_s$ does not depend on spin. Summing over both spin indices, we
obtain with a partial wave decomposition,
\bwt \be
\label{eq:gSpinSum}
g(\R)=\frac{\sum_{\k,m}\exp\left(-\frac{\hbar^2k^2}{Mk_BT}\right)
\left\{3\abs{\psi_m^a(R)}^2+\abs{\psi_m^s(R)}^2
+(-1)^m\left(\frac{3}{2}\abs{\psi_m^a(R)}^2-\frac{1}{2}
\abs{\psi_m^s(R)}^2\right)\right\}}{4\sum_{\k}\exp\left(
-\frac{\hbar^2k^2}{Mk_BT}\right)}.
\ee \ewt
The results shown in \Fig{fig:fig11_Pair_correlation_Butov} reflect the strong
repulsion of the excitons independent of the spin channel. It is
interesting to compare with the pair-correlation function of
ideal bosons and fermions having four spin degrees of freedom as
well. Obviously in the present case, the repulsive interaction
between excitons is much more important than the bosonic nature
of their statistics.

\section{Conclusions}
\label{sec:conclusions} In the investigation of the many-exciton
problem in semiconductor quantum structures, we have found a much
richer spin structure with spatially dependent interaction
potentials than with just contact interactions. This is due to
non trivial exchange processes of the fermionic constituents. For
the spatially dependence of the potentials, we compared three
different levels of approximation: The Hartree-Fock and the
Heitler-London approximations as well as a newly introduced full
numerical solution of the two exciton problem. We found a
principal failure of the Hartree-Fock treatment, which is cured
in the Heitler-London approach. The quality of the latter,
however, turns out to be quite poor, compared to numerically
exact results. With our calculated potentials we have
investigated bilayer systems and two different CQWs. The charge
separation $d$ plays a fundamental role: By tuning $d$ a
transition happens from systems with possible biexciton formation
to those where biexcitons are not bound due to the stronger XX
repulsion. For two realistic CQW structures we have calculated
the quasiparticle shift and broadening at the band edge which
govern roughly the photoluminescence line shape. At low
temperatures, we found a dramatic reduction of the blueshift
compared to a naive treatment of the CQW as plate capacitor. The
broadening turns out to be of the same order as the blueshift.

\acknowledgments
Numerous stimulating discussions with Leonid Butov, Boris
Laikhtman, Vincenzo Savona, and Dave Snoke are gratefully
acknowledged.

\appendix
\bwt
\section{ }\label{app:HXX}
In the limit of infinitely heavy holes, the interaction part of
the many-exciton Hamiltonian (Table \ref{tab:tbl2_W_0_spin_struct}) can
be written as\cite{Zim4}
\begin{multline} \label{eq:ManyXHamInt}
H_{XX}^{\text{int}}= \frac{1}{2}\int d\R d\R' U^a(\R-\R')\sum_{ss'}
\Psi_s^{\dagger}(\R) \Psi_{s'}^{\dagger}(\R') \Psi_{s'}(\R')\Psi_s(\R)\\
 + \frac{1}{4}\int d\R d\R' \left( U^a(\R-\R')-U^s(\R-\R')\right)\sum_{ss'}
 \Theta(s\cdot s')\\
 \left[ \Psi_{s'}^{\dagger}(\R)\Psi_{s}^{\dagger}(\R') \Psi_{s'}(\R')
 \Psi_{s}(\R) - \Psi_{s}^{\dagger}(\R)\Psi_{s'}^{\dagger}(\R') \Psi_{s'}(\R')
 \Psi_{s}(\R)\right.\\
 \left.+ \Psi_{s'}^{\dagger}(\R)\Psi_{-s'}^{\dagger}(\R') \Psi_{-s}(\R')
 \Psi_{s}(\R)\left( 1- 2\delta_{ss'}\right) \right].
\end{multline}
For the simplifications of contact potentials
$U^p(\R-\R')=U^p\delta(\R-\R')$, our result agrees with the one
derived by de-Leon and Laikhtman.\cite{Laikhtman1} This can be seen as
follows: Plugging the contact potentials into \Eq{eq:ManyXHamInt}
the integration over $\R'$ can be carried out immediately,
\begin{multline}
H_{XX}^{\text{int}}= \frac{1}{2}\,U^{a}\int d\R \sum_{ss'}\Psi_s^{\dagger}(\R)
\Psi_{s'}^{\dagger}(\R) \Psi_{s'}(\R)\Psi_s(\R)\\
 + \frac{1}{4}\left( U^a-U^s\right)\int d\R\sum_{ss'}\Theta(s\cdot s')
 \Psi_{s'}^{\dagger}(\R)\Psi_{-s'}^{\dagger}(\R) \Psi_{-s}(\R)
 \Psi_{s}(\R)\left( 1- 2\delta_{ss'}\right).
\end{multline}
In this limit, the third row of \Eq{eq:ManyXHamInt} gives no
contribution.  We can also drop the Heaviside step function by
inserting a factor of $1/2$ to account for the double counting in
the sum over $s$ and $s'$ and get
\begin{multline} \label{eq:A3}
H_{XX}^{\text{int}}= \frac{1}{2}\,U^{a}\int d\R \sum_{ss'}\Psi_s^{\dagger}(\R)
\Psi_{s'}^{\dagger}(\R) \Psi_{s'}(\R)\Psi_s(\R)\\
 + \frac{1}{8}\left( U^a-U^s\right)\int d\R\sum_{ss'} \Psi_{s'}^{\dagger}(\R)
 \Psi_{-s'}^{\dagger}(\R) \Psi_{-s}(\R)\Psi_{s}(\R)
 \left( 1- 2\delta_{ss'}\right).
\end{multline}
In the language of the Hartree-Fock approximation, we can trade
the symmetric and the antisymmetric potentials for the direct and
the exchange ones. For arbitrary hole masses, $U_d(\R)$ and
$U_x(\R)$ have been derived in the literature.\cite{Brandes1} For
immobile holes they reduce to the direct potential,
\be \label{eq:Ud} U_d(\R)=v_{hh}(\R)+\int d\r d\r' \phi^2(\r)
v_{ee}(\r-\r') \phi^2(\r'-\R)-2\int d\r v_{eh}(\r) \phi^2(\r-\R),
\ee
and the exchange potential,
\be \label{eq:Ux} U_x(\R)=-\mathcal{O}^2(\R)v_{hh}(\R) -\int d\r
d\r' \phi(\r) \phi(\r-\R) v_{ee}(\r-\r') \phi(\r') \phi(\r'-\R)
+2\mathcal{O}(\R)\int d\r v_{eh}(\r) \phi(\r) \phi(\r-\R). 
\ee
Taken in the contact limit $U^a=U_d+U_x$ and $U^s=U_d-U_x$,
\Eq{eq:A3} can be written as
\begin{multline}
H_{XX}^{\text{int}}= \frac{1}{2}\,U_d\int d\R \sum_{ss'}\Psi_s^{\dagger}(\R)
\Psi_{s'}^{\dagger}(\R) \Psi_{s'}(\R)\Psi_s(\R)\\
 + \frac{1}{4}\,U_x\int d\R\sum_{ss'}\left[ 2\,\Psi_s^{\dagger}(\R) 
 \Psi_{s'}^{\dagger}(\R) \Psi_{s'}(\R)\Psi_s(\R)\right.
 \left.+ \Psi_{s'}^{\dagger}(\R)\Psi_{-s'}^{\dagger}(\R) \Psi_{-s}(\R)
 \Psi_{s}(\R)\left( 1- 2\delta_{ss'}\right) \right].
\end{multline}
The Fourier transform of this result yields the Hamiltonian
 derived in \Quot{Laikhtman1}.
~\\
\ewt

\bibliography{Schindler_XX}
\end{document}